\documentclass[aps,twocolumn,prb,floatfix,longbibliography,superscriptaddress,amssymb,amsmath]{revtex4-2}
\pdfoutput=1
\usepackage{graphicx,braket,float}
\usepackage{amsmath}
\usepackage{amsfonts}
\usepackage{amsbsy}
\usepackage{xcolor}
\usepackage{soul}
\usepackage{bm}
\usepackage[normalem]{ulem}
\usepackage[unicode]{hyperref}
\usepackage[export]{adjustbox}
\usepackage{enumitem}
\hypersetup{
    unicode=true, 
    plainpages=false,
    colorlinks=true,
    linkcolor=blue,
    citecolor=blue,
    filecolor=blue,
    urlcolor=blue
}
\newcommand{\Tr}{\operatorname{Tr}}

\begin{document}

\title{Many-body enhancement in a spin-chain quantum heat engine}
\author{L. A. Williamson}
\affiliation{ARC Centre of Excellence for Engineered Quantum Systems, School of Mathematics and Physics, University of Queensland, St Lucia, Queensland 4072, Australia}
\author{Matthew J. Davis}
\affiliation{ARC Centre of Excellence for Engineered Quantum Systems, School of Mathematics and Physics, University of Queensland, St Lucia, Queensland 4072, Australia}
\date{\today}

\begin{abstract}
We show that ferromagnetic interactions can enhance the adiabatic performance of a quantum spin chain engine at low temperatures. The enhancement in work output is particular pronounced, increasing exponentially with interaction strength. The performance enhancement occurs in the paramagnetic phase and is qualitatively explained by considering just the ground and first excited state, in which case the system exhibits bipartite entanglement. As the temperature is increased, thermal occupation of higher energy states diminishes performance. We find that these thermal fluctuations are smallest for long-range interactions, resulting in the highest efficiency. Diabatic work extraction degrades performance due to quantum friction. We identify an approximate, experimentally realisable counterdiabatic drive that can mitigate friction for weak interactions.
\end{abstract}

\maketitle

\section{Introduction}

Quantum heat engines convert heat into work utilising some distinctly quantum effect in the reservoir or working substance~\cite{millen2016}. Reservoirs possessing coherence~\cite{scully2003,hardal2015,hammam2022}, squeezing~\cite{huang2012,robnagel2014,manzano2016,niedenzu2016,klaers2017,agarwalla2017,niedenzu2018} or entanglement~\cite{dillenschneider2009,obah2014} have been shown to improve engine performance. Coherence in a working substance can be utilised as a resource~\cite{kammerlander2016,korzekwa2016,uzdin2016,tajima2021} and can improve the power output of rapid engine cycles~\cite{uzdin2015,klatzow2019}. In the many-body regime, interactions in a Bose gas can enhance engine performance compared to a non-interacting gas~\cite{bengtsson2018,fogarty2021,boubakour2023}. Interactions in a many-body quantum system can also be tuned to change the energy of a working substance, hence providing a means to extract work~\cite{chen2019,halpern2019,carollo2020}.

One of the simplest quantum working substances is an ensemble of two-level systems (``spins'')~\cite{geva1992,vonlindenfels2019,peterson2019,klatzow2019,niedenzu2018b,kloc2019,ono2020,wang2012b,wang2013,kieu2004,ji2022}. Work can be extracted by tuning the level spacing $\hbar\omega(t)$ via the control of an external field, see Fig.~\ref{fig2}(a). Including interactions between spins opens up the possibility to explore many-body quantum effects. While considerable work has explored engines with two interacting spins~\cite{kosloff2002,feldmann2003,feldmann2004,zhang2007,zhang2008,wang2009,thomas2011,altintas2014,altintas2015,mehta2017,cakmak2016,zhao2017,hewgill2018,chand2018}, much less is known about greater numbers of spins. For nearest-neighbour interactions, a spin chain can function as both a heat engine and a refrigerator~\cite{piccitto2022} with critical scaling of performance close to the critical point~\cite{revathy2020}. While moderate enhancements due to interactions have been identified in systems of two spins~\cite{thomas2011,altintas2014,altintas2015,mehta2017}, a thorough investigation of whether many-body effects can improve the performance of a spin-chain quantum heat engine is lacking.

In this work we characterise the performance of an Otto cycle with a ferromagnetic spin chain as the working substance. In addition to displaying rich many-body physics, this system may be realised in experiments with a remarkable degree of control~\cite{porras2004,britton2012,bohnet2016,zhang2017,monroe2021}. We show that both short and long-range interactions improve the adiabatic work output and efficiency in the paramagnetic phase at low temperatures $k_B T\lesssim \hbar\omega$. The performance enhancement is qualitatively explained by an analytic model considering just the ground and first excited state, in which case the thermal state exhibits bipartite entanglement. For temperatures $k_BT>\hbar\omega$, higher energy eigenstates are occupied and interactions degrade performance. These thermal fluctuations decrease as the range of interactions is increased, and hence greater efficiency is most robust for long-range interactions. For diabatic work extraction, decreasing the engine cycle time reduces performance due to quantum friction~\cite{kosloff2002,feldmann2003,plastina2014}. We demonstrate an approximate, experimentally realisable counterdiabatic drive that can mitigate friction for weak interactions, and hence a performance enhancement is possible at finite power output.

This paper is organised as follows. In Sec.~\ref{model} we introduce the spin-chain model and the engine cycle. In Sec.~\ref{adiabaticPerformance} we demonstrate the performance enhancement in the adiabatic, low temperature limit. In Sec.~\ref{finiteTempPerformance} we demonstrate how increasing temperature decreases the performance enhancement. In Sec.~\ref{diabaticPerformance} we show performance for finite-time work extraction, and show how a performance enhancement can be retained by using an approximate counter-diabatic drive. We conclude in Sec.~\ref{conclusion}.

\section{Model}\label{model}
A chain of $N$ ferromagnetic interacting two-level spins is described by the Hamiltonian (hereon $\hbar\equiv 1$),
\begin{equation}\label{H}
\hat{H}(\omega(t))=-\omega(t)\sum_{i=1}^N\hat{\sigma}_z^{(i)}-g\sum_{\substack{i,j=1\\(j\ne i)}}^NJ_{ij}\hat{\sigma}_x^{(i)}\hat{\sigma}_x^{(j)},
\end{equation}
with $\hat{\sigma}_\mu^{(i)}$ ($\mu=x,y,z$) the Pauli spin-1/2 matrices for spins $i=1,...,N$. The interaction strength between spins $i$ and $j$ is $g J_{ij}$ with $J_{ij}=1/|i-j|^p$, $g\geq 0$ the nearest-neighbour interaction strength and $p>0$ determining the range of interactions. Both $g$ and $p$ are tuneable in experiments~\cite{britton2012,zhang2017}. For $N\rightarrow \infty$, the system may be paramagnetic $(g\lesssim \omega)$ or ferromagnetic $(g\gtrsim\omega)$ with the precise cross-over $g_c(p)$ dependent on $p$~\cite{dutta2001,fey2016,defenu2017,vanderstraeten2018,zhu2018,gonzalezlazo2021,koziol2021}. We denote nearest-neighbour interactions by $p=\infty$.

We consider an Otto engine cycle with the following steps, as shown in Fig.~\ref{fig2}(a).
\begin{itemize}[leftmargin=1.3cm]
\item [(1)] We begin with a hot thermal state $\rho^\mathrm{th}_H=e^{-\beta_H\hat{H}(r\omega_0)}/Z(\beta_H,r\omega_0)$ at level spacing $\omega=r\omega_0$, with $r>1$ the ``compression ratio'' (in analogy with the ideal gas Otto cycle, where the compression ratio is given by the ratio of maximum and minimum gas volumes). The partition function is $Z(\beta,\omega)=\Tr e^{-\beta\hat{H}(\omega)}$ and $\beta=(k_B T)^{-1}$ is the inverse temperature.
\item [(1 $\rightarrow$ 2)] The system is then thermally isolated and work is extracted by decreasing $\omega$ from $r\omega_0$ to $\omega_0$, via the protocol $\omega(t)/\omega_0=f(t)$ ($0\le t\le\tau$). We choose $f(t)=r+(1-r)\sin^2(\pi t/2\tau)$.
\item [(2 $\rightarrow$ 3)] Next, we cool the system at fixed $\omega=\omega_0$, leaving the system in a cold thermal state $\rho^\mathrm{th}_C=e^{-\beta_C\hat{H}(\omega_0)}/Z(\beta_C,\omega_0)$.
\item [(3 $\rightarrow$ 4)] We thermally isolate the system again and increase $\omega$ from $\omega_0$ back to $r\omega_0$, with the protocol $\omega(t)/\omega_0=f(\tau-t)$ ($0\le t\le \tau$).
\item [(4 $\rightarrow$ 1)] Finally we heat the system at fixed $\omega=r\omega_0$ back to the initial state.
\end{itemize}

The work output $W$ and efficiency $\eta$ of the engine cycle are
\begin{equation}
W=Q_H-Q_C,\hspace{0.6cm}\eta=\frac{W}{Q_H}.
\end{equation}
Here $Q_H=\Tr\left[\hat{H}(r\omega_0)(\rho^\mathrm{th}_H-\rho_4)\right]$ is the heat input from the hot reservoir and $Q_C=-\Tr\left[\hat{H}(\omega_0)(\rho^\mathrm{th}_C-\rho_2)\right]$ is the heat output to the cold reservoir, with $\rho_4$ the density matrix prior to coupling to the hot reservoir and $\rho_2$ the density matrix prior to coupling to the cold reservoir. The density matrix at points 2 and 4 are obtained by time-evolving the von Neumann equation $\dot{\rho}(t)=-i[H(t),\rho(t)]$ with initial conditions $\rho^\mathrm{th}_H$ and $\rho^\mathrm{th}_C$ respectively, using Runge-Kutta integration.

\section{Adiabatic low-temperature performance}\label{adiabaticPerformance}

\subsection{Interactions enhance performance}
\begin{figure}
\includegraphics[trim=0cm 2cm 0cm 2cm,clip=true,width=0.5\textwidth]{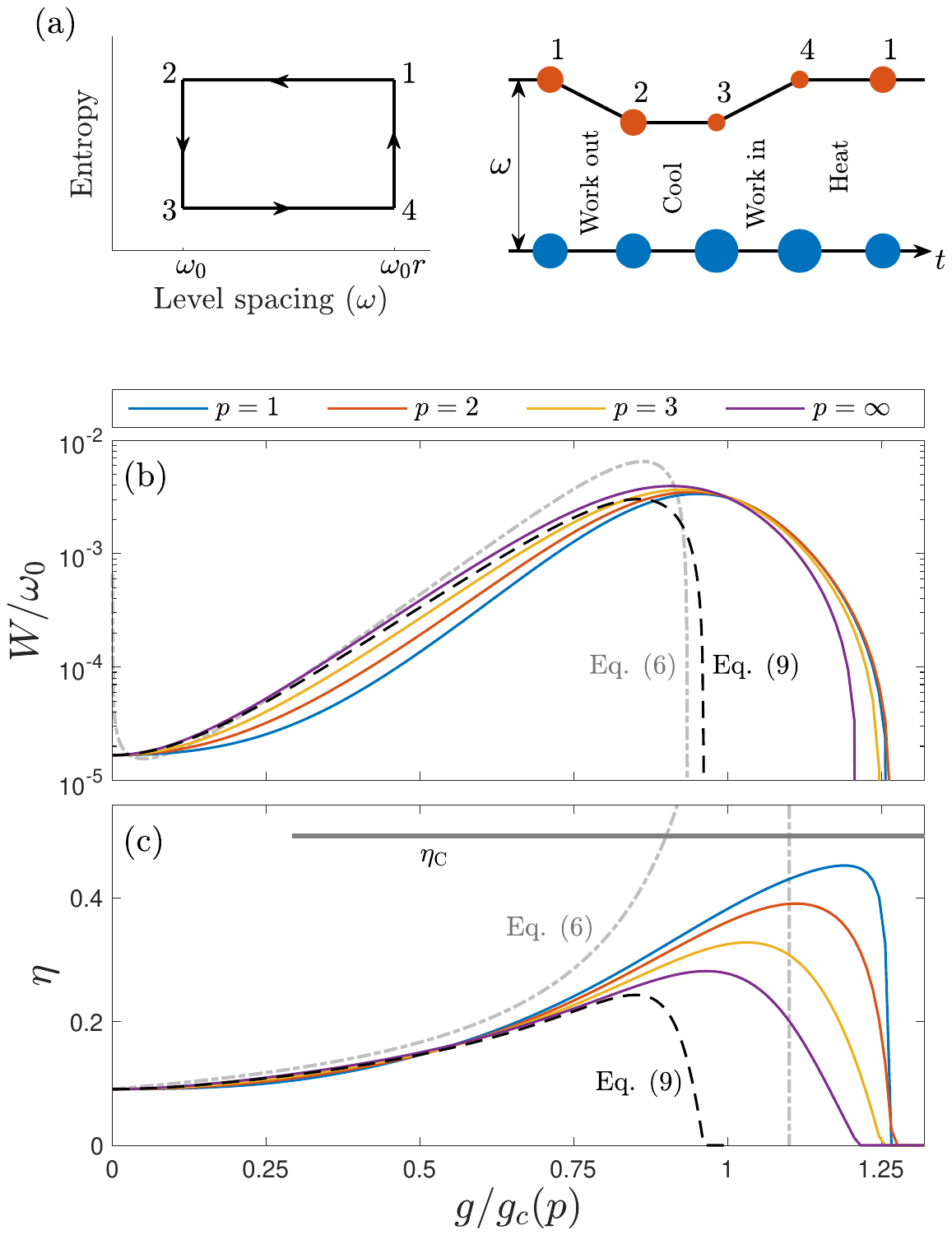}
\caption{\label{fig2} (a) An Otto cycle can be realised in an ensemble of two-level spins, as described in the main text. The area of blue and red spots indicate the relative ground and excited state occupations, respectively. (b),(c) Engine performance for a spin chain operating adiabatically at low temperature ($\beta_H=10\omega_0^{-1}$, $\beta_C=2\beta_H$, $N=10$). Increasing $g$ increases both (b) work output and (c) efficiency, with maximum performance at $g\approx g_c(p)$ (results are for $r=r_\mathrm{NI}^\mathrm{max}$). Above this, the system transitions to the ferromagnetic phase and no longer functions as an engine. The approximate $p=\infty$ performance computed from Eq.~\eqref{2level} (gray dot-dashed line) and from Eq.~\eqref{EIsing} (black dashed line) are also shown. The gray horizontal line in (c) is the Carnot efficiency.}
\end{figure}

We first examine the quantum adiabatic limit $\tau\gg \omega_0^{-1},g^{-1}$ (we set $\tau=100\omega_0^{-1}$) such that transitions between eigenstates during the work steps are suppressed~\cite{born1928,kato1950,avron1999,griffiths2018}. For zero interactions and fixed $\beta_H\omega_0\gg 1$, the maximum work output occurs at a compression ratio $r_\mathrm{NI}^\mathrm{max}\approx 1+(\beta_H\omega_0)^{-1}$, which gives a small efficiency $\eta_\mathrm{NI}\approx (\beta_H\omega_0)^{-1}$ that decreases with decreasing temperature. We find that interactions drastically improve both work output and efficiency in the paramagnetic phase for temperatures $\beta_H^{-1}\ll\omega_0$, see Fig.~\ref{fig2}(b),(c). At these low temperatures, the performance enhancement is far greater than that identified in systems of two interacting spins~\cite{thomas2011,altintas2014,altintas2015,mehta2017}. The improvement in work output is particularly pronounced, with a maximum work output $\sim 10^2$ times larger than the non-interacting ensemble. The behaviour is qualitatively similar in all cases $p=1,2,3,\infty$ after rescaling interactions by $g_c(p)$, which we define to be the point at which $\partial^2\Delta/\partial g^2\big|_{\omega=\omega_0}$ has a maximum (see Appendix~\ref{gccalc}). Here $\Delta$ is the energy gap to the first excited state. The improvement increases monotonically up to $g\approx g_c(p)$, before dropping abruptly.

We now derive an approximate analytic theory that quantitatively describes the performance enhancement for low spin excitation and qualitatively describes the performance enhancement generally. The spin operators can be written in terms of bosonic operators via a Holstein-Primakoff transformation~\cite{holstein1940}. Expanding to quadratic order in bosonic operators gives an analytically tractable theory. To lowest order in $g/\omega$ and for large $N$ and $\beta$ we obtain (see Appendix~\ref{perturb}),
\begin{equation}\label{Zquadratic}
\ln Z\approx N\mathcal{G}_p(\beta g)e^{-\beta \Delta},
\end{equation}
where $Ne^{-\beta\Delta}$ is the low-temperature free energy of $N$ two-level systems with level splitting $\Delta(\omega)=\omega-\omega_0g/g_c$. We have assumed a frame where the ground-state energy is zero. The factor $\mathcal{G}_p(\beta g)$ arises from thermal fluctuations and depends on $p$,
\begin{equation}
\mathcal{G}_p(\beta g)\approx \left\{\begin{array}{ll}\sqrt{\frac{1}{2\pi\beta g\zeta(p-2)}},&\hspace{0.5cm}p>3,\\
\sqrt{\frac{1}{\pi\beta g(3+\ln\beta g)}},&\hspace{0.5cm}p=3,\\
\frac{1}{3\beta g\zeta(2)},&\hspace{0.5cm}p=2,\\
1,&\hspace{0.5cm}p=1,\end{array}\right.
\end{equation}
with $\zeta(s)=\sum_{m=1}^\infty 1/m^s$ the Riemann zeta function.

Using Eq.~\eqref{Zquadratic}, the system energy $\langle E_i\rangle$ at the four points $i=1,2,3,4$ in the cycle in Fig.~\ref{fig2}(a) under adiabatic operation are,
\begin{equation}\label{energies}
\begin{split}
\langle E_1\rangle&=N\Delta(r\omega_0)\mathcal{G}_p(\beta_H g)e^{-\beta_H\Delta(r\omega_0)},\\
\langle E_2\rangle&=N\Delta(\omega_0)\mathcal{G}_p(\beta_H g)e^{-\beta_H\Delta(r\omega_0)},\\
\langle E_3\rangle&=N\Delta(\omega_0)\mathcal{G}_p(\beta_C g)e^{-\beta_C\Delta(\omega_0)},\\
\langle E_4\rangle&=N\Delta(r\omega_0)\mathcal{G}_p(\beta_C g)e^{-\beta_C\Delta(\omega_0)}.\\
\end{split}
\end{equation}
We neglect a correction term $\partial\mathcal{G}_p(\beta g)/\partial\beta=O(\beta^{-1}\mathcal{G}_p(\beta g))$ in Eq.~\eqref{energies}, which is valid for low temperatures. Equation~\eqref{energies} gives,
\begin{equation}\label{2level}
\begin{split}
W=&N\omega_0(r-1)\Big[\mathcal{G}_p(\beta_Hg)e^{-\beta_H\Delta(r\omega_0)}\\
&\hspace{2cm}-\mathcal{G}_p(\beta_Cg)e^{-\beta_C\Delta(\omega_0)}\Big],\\
\eta=&1-\frac{\Delta(\omega_0)}{\Delta(r\omega_0)}.
\end{split}
\end{equation}
For $g<g_c(p)$, increasing $g$ decreases $\Delta(\omega)$. From examination of Eq.~\eqref{2level}, this increases low-temperature work output as $W\sim W_\mathrm{NI}e^{\beta_H\omega_0 g/g_c}$, consistent with the exponential increase in Fig.~\ref{fig2}(b), and efficiency as $\eta\sim \eta_\mathrm{NI}/(1-g/rg_c)$. Above $g_c$ the system transitions to the ferromagnetic state and $\Delta$, and therefore $\partial^2\ln Z/\partial\beta\partial\omega$, changes sign. The cycle therefore no longer functions as a heat engine~\cite{piccitto2022}, resulting in the abrupt drop in performance above $g_c$ in Fig.~\ref{fig2}(b),(c). The analytic result Eq.~\eqref{2level} for $p=\infty$ is shown in Fig.~\ref{fig2}(b),(c).

The bosonic approximation above permits a calculation of the bipartite entanglement of the spin chain. For low temperatures, a thermal state can be approximated by $\rho_1\approx(\ket{0}\bra{0}+e^{-\beta\Delta}\ket{1}\bra{1})/(1+e^{-\beta\Delta})$, with $\ket{0}$ the ground state, $\ket{1}=\sum_{i=1}^N\hat{\sigma}_+^{(i)}\ket{0}/\sqrt{N}$ the approximate first-excited state (independent of $p$) and $\hat{\sigma}_+^{(i)}=\hat{\sigma}_x^{(i)}+i\hat{\sigma}_y^{(i)}$. In Appendix~\ref{entangle}, we show that $\rho_1$ is entangled according to the Peres-Horodecki criterion~\cite{peres1996,horodecki1996,toth2009}. The performance enhancement, Eq.~\eqref{2level}, requires access to an entangled thermal state, and so coincides with many-body quantum effects. The entanglement of $\ket{1}$ is also directly evident from the entanglement entropy, which is $\mathcal{S}=\ln(N/2)$ for a partition dividing the spin chain in half, see Appendix~\ref{entangle}. In contrast, in a mean-field approximation, the interaction of spin $i$ with the remaining spins is replaced by $-g\Omega_i\hat{\sigma}_x^{(i)}$. Here $\Omega_i=2\sum_{j\ne i}J_{ij}s_j$ is an effective transverse drive and $s_j=\langle\hat{\sigma}_x^{(j)}\rangle_\mathrm{mf}$ is a mean-field approximation for spin $j$. The energy gap of spin $i$ then increases with $g$ as $\sqrt{\omega^2+g^2\Omega_i^2}$ and interactions degrade performance.

In the limit of large $N$, a chain with nearest neighbour interactions maps onto a system of non-interacting fermions with energies~\cite{pfeuty1970}
\begin{equation}
\mathcal{E}_\omega(\theta)=\sqrt{\omega^2+g^2-2\omega g\cos \theta}.
\end{equation}
For $\beta_H^{-1}\ll\omega_0,g$, only the lowest energy states are appreciably occupied. Considering only the ground state and eigenstates with singly occupied fermions gives
\begin{equation}\label{ZIsing}
\ln Z\approx \frac{N}{\pi}\int_0^{\pi}e^{-\beta\mathcal{E}_\omega(\theta)}\,d\theta.
\end{equation}
In the adiabatic limit, the system energies $\langle E_i\rangle$ computed from Eq.~\eqref{ZIsing} are
\begin{equation}\label{EIsing}
\begin{split}
\langle E_1\rangle&=\frac{N}{\pi}\int_0^\pi \mathcal{E}_{r\omega_0}(\theta)e^{-\beta_H\mathcal{E}_{r\omega_0}(\theta)}\,d\theta,\\
\langle E_2\rangle&=\frac{N}{\pi}\int_0^\pi \mathcal{E}_{\omega_0}(\theta)e^{-\beta_H\mathcal{E}_{r\omega_0}(\theta)}\,d\theta,\\
\langle E_3\rangle&=\frac{N}{\pi}\int_0^\pi \mathcal{E}_{\omega_0}(\theta)e^{-\beta_C\mathcal{E}_{\omega_0}(\theta)}\,d\theta,\\
\langle E_4\rangle&=\frac{N}{\pi}\int_0^\pi \mathcal{E}_{r\omega_0}(\theta)e^{-\beta_C\mathcal{E}_{\omega_0}(\theta)}\,d\theta.
\end{split}
\end{equation}
The performance computed from Eq.~\eqref{EIsing} is plotted alongside the full numerical results in Fig.~\ref{fig2}(b),(c) and agrees well with the $p=\infty$ results for $g<g_c$.

\subsection{Effect of compression ratio}\label{compression}
\begin{figure}
\includegraphics[trim=0cm 6cm 0cm 6cm,clip=true,width=0.5\textwidth]{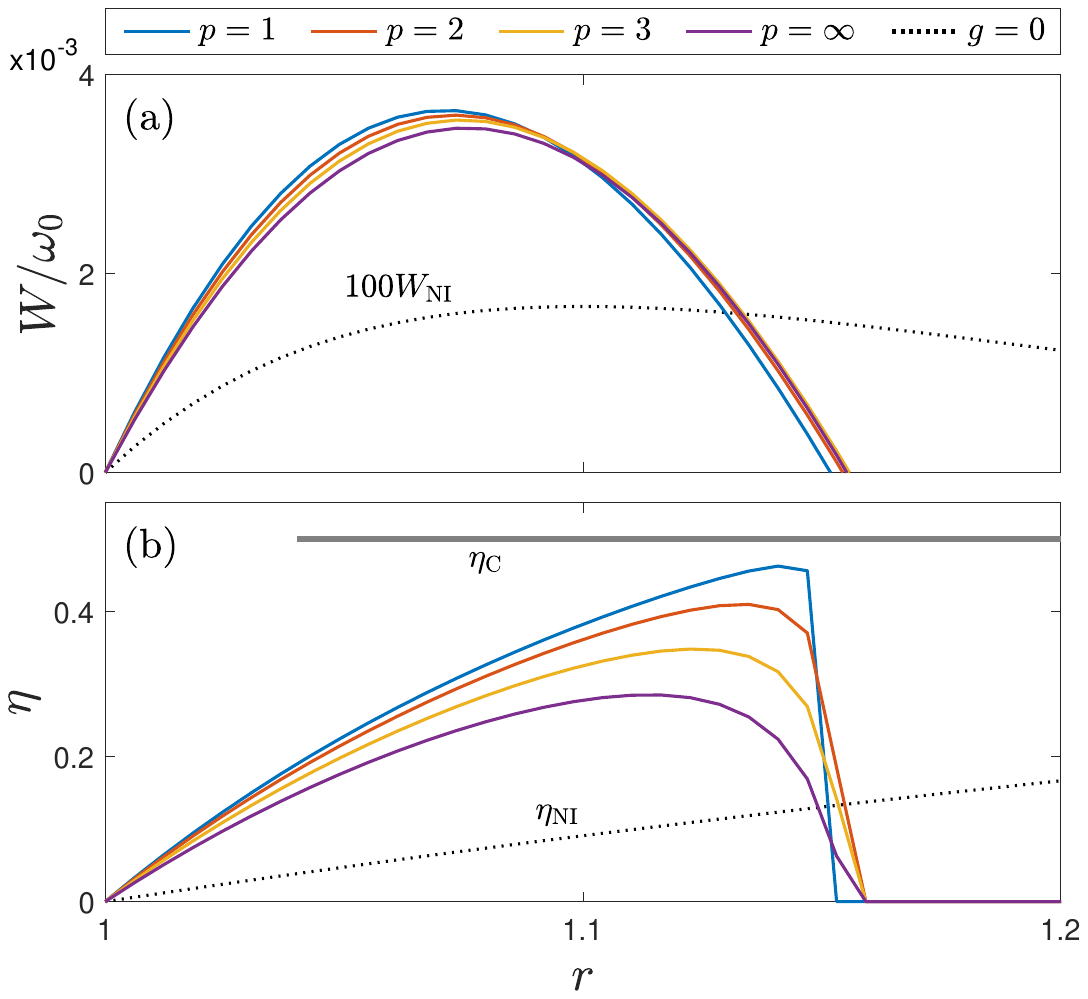}
\caption{\label{fig3} Engine performance for a spin chain operating adiabatically at low temperature ($\beta_H=10\omega_0^{-1}$, $\beta_C=2\beta_H$, $N=10$). (a) Work output and (b) efficiency as a function of compression ratio for $g=g_c(p)$, with dotted lines indicating the non-interacting results $W_\mathrm{NI}$ and $\eta_\mathrm{NI}$ ($W_\mathrm{NI}$ is scaled by a factor of $100$). The gray horizontal line in (b) is the Carnot efficiency.}
\end{figure}

Increasing the compression ratio increases performance until $r=r^\prime$, with $r^\prime\sim 1.1$ at $g=g_c(p)$. For larger compression ratio the performance abruptly drops, see Fig.~\ref{fig3}. Unlike the non-interacting case, the peak work output and efficiency can both occur at a comparable compression ratio. Equation~\eqref{2level} describes this behaviour: within this approximation, performance increases until $1-\Delta(\omega_0)/\Delta(r\omega_0)\sim \eta_C$, with $\eta_C=1-\beta_H/\beta_C\equiv 1-1/r_C$ the Carnot efficiency. Hence $r^\prime\approx r_C-(r_C-1)g/g_c+O(g^2)$ decreases with increasing $g$. As a result, for $g\sim g_c(p)$ we can have high efficiency at small compression ratios $r\sim r_\mathrm{NI}^\mathrm{max}$. Without interactions, the efficiency is maximum for a compression ratio $r_C\gg r_\mathrm{NI}^\mathrm{max}$.

Choosing $\omega_0$ and $r$ to maximise the work output in Eq.~\eqref{2level} gives the same efficiency as a cold two-level atom at maximum work output (ignoring corrections due to $\mathcal{G}_p$). This is approximately the Curzon-Ahlborn efficiency~\cite{allahverdyan2008}. The value of $\omega_0$ obtained satisifies $(1-g/g_c)\beta_C\omega_0\approx 1+\eta_C^{-1}\ln r_C$. For $\eta_C^{-1}\ln r_C\sim 1$ and $\beta_C\omega_0\gg 1$, we are close to this regime only if $g\approx g_c$.

\subsection{Finite-size scaling}
\begin{figure}[H]
\includegraphics[trim=0cm 1cm 0cm 1cm,clip=true,width=0.44\textwidth,center]{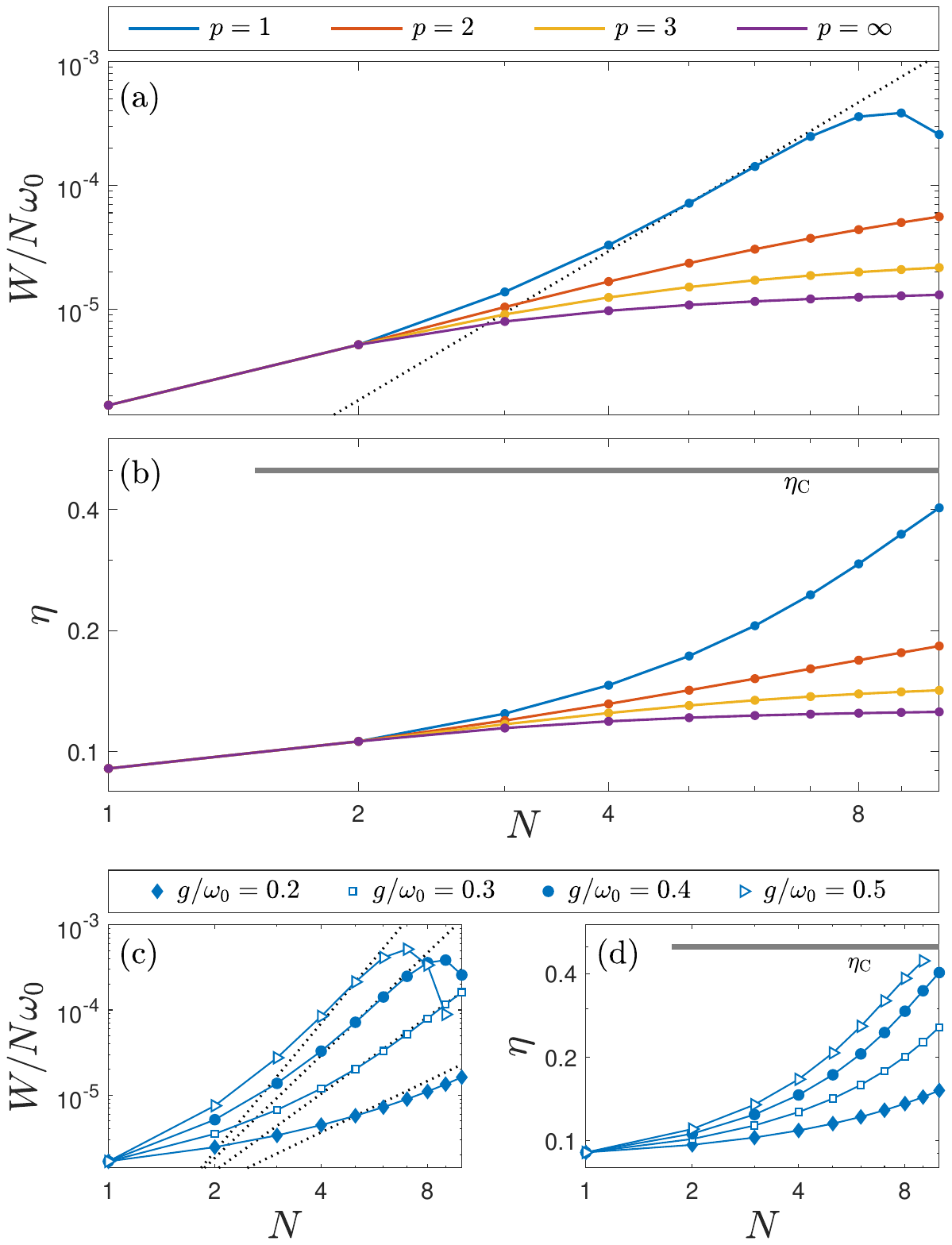}
\caption{\label{suppfig2} The (a) work output and (b) efficiency for increasing system size for $g=0.4\omega_0$, showing non-extensive scaling for $p=1$. The perturbative prediction $W\sim N^{1+\beta_H g}$ is also shown (black dotted line). (c),(d) $p=1$ performance for additional values of $g/\omega_0$ (blue symbols) along with the perturbative predictions for $W$ (matching black dotted lines). The gray lines in (b),(d) are the Carnot efficiency. All results are for $\beta_H=10\omega_0^{-1}$, $\beta_C=2\beta_H$ and $r=r_\mathrm{NI}^\mathrm{max}$.}
\end{figure}

The dependence of work and efficiency on chain size $N$ is shown in Fig.~\ref{suppfig2}. For $p=\infty$ and $p=3$ the performance tends toward extensive scaling as $N$ is increased. Finite-size effects are slightly more pronounced for $p=2$; however, the scaling is close to extensive. For $p=1$ the performance increases non-extensively due to a dependence of $\Delta$ on $N$ (see inset to Fig.~\ref{suppfig1}(a) in Appendix~\ref{gccalc}). Non-extensive thermodynamics is expected in systems with long-range interactions~\cite{campa2009,defenu2023}. The critical point $g_c$ also depends on $N$ for $p=1$. In the cold, perturbative limit (see Appendix~\ref{perturb}), $\Delta\sim \omega-g(\gamma+\ln N)$ with $\gamma$ the Euler-Mascheroni constant. The two-level approximation, Eq.~\eqref{2level}, then gives $W\sim N^{1+\beta_H g}$. This approximation captures the dependence of work on particle number for small $g/g_c$, see Fig.~\ref{suppfig2}(c). Increasing either $g$ or $N$ increases $g/g_c$ and the perturbative approximation $W\sim N^{1+\beta_H g}$ breaks down. Scaling $g$ by $g_c$ effectively Kac renormalizes the interactions, rendering the long-range system extensive~\cite{kac1963,kac1969,cannas1996,gonzalezlazo2021,zhang2017}.

\section{Adiabatic performance for increasing temperature}\label{finiteTempPerformance}

\begin{figure}
\includegraphics[trim=0cm 2cm 0cm 2cm,clip=true,width=0.5\textwidth]{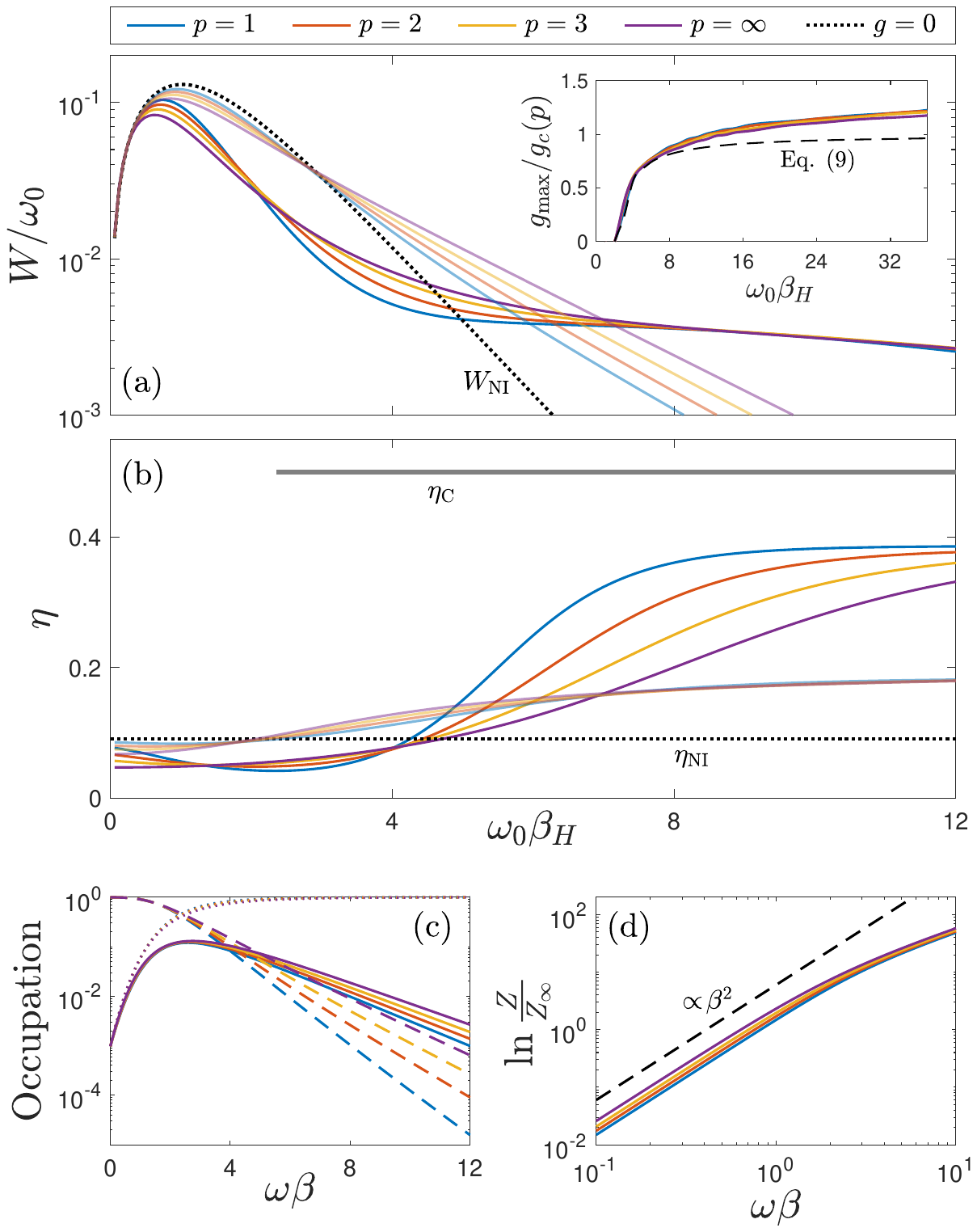}
\caption{\label{fig4} Adiabatic engine performance for varying temperature. (a) Work output and (b) efficiency exceed the non-interacting values (dotted lines) for $\beta_H\gtrsim 4\omega_0^{-1}$ [dark lines: $g=g_c(p)$; light lines: $g=0.8g_c(p)$; $r=1.1$]. The gray line in (b) is the Carnot efficiency. Inset in (a) shows dependence of $g_\mathrm{max}/g_c(p)$ on $\beta_H$ with $r=r_\mathrm{NI}^\mathrm{max}$; dashed line is $g_\mathrm{max}/g_c(p)$ computed from Eq.~\eqref{EIsing}. (c) Thermal energy-level occupations $n_0$ (dotted lines), $n_1$ (solid lines) and $\sum_{i=2}^{2^N-1} n_i$ (dashed lines) at $g=g_c(p)$. The performance enhancement occurs when occupation is predominantly in the ground and first excited state. Long-range interactions suppress occupation beyond the first excited state, resulting in the highest efficiency in (b). (d) Dimensionless free energy $\ln Z$ at $g=g_c(p)$ showing the $\beta^2$ scaling (dashed line) for $\beta\lesssim \omega^{-1}$, coinciding with the regime of reduced performance. All results are for $N=10$ with $\beta_C=2\beta_H$.}
\end{figure}

As the temperature increases, thermal fluctuations render Eq.~\eqref{2level} invalid and we find that the performance enhancement relative to the non-interacting system is diminished, see Fig.~\ref{fig4}(a),(b). A performance enhancement is present as long as $\beta_H\gtrsim 4\omega_0^{-1}$, coinciding with the regime where only the ground and first excited state are appreciably occupied, see Fig.~\ref{fig4}(c). The interaction strength $g_\mathrm{max}$ that gives maximum work output increases toward $g_c(p)$ for decreasing temperature, consistent with the discussion in Sec.~\ref{compression}, see inset to Fig.~\ref{fig4}(a). Curiously, we see a gradual increase in $g_\mathrm{max}$ above $g_c(p)$ for very low temperatures. This is not accounted for by Eq.~\eqref{EIsing}, and so we expect is due to finite-size effects.

The transverse Ising model gives a qualitative understanding of the diminished performance enhancement at higher temperatures. The energy levels of this model are sums of fermion energies $\mathcal{E}_\omega(\theta_k)$, with $\theta_k=2\pi k/N$ ($k=0,...,N-1$)~\cite{pfeuty1970}. Interactions diminish fermion energies with $\cos\theta_k>g/\omega$, with the most pronounced reduction occurring for the lowest energy fermion ($k=0$). Hence, the enhancement is largest when only the first excited state is occupied, and diminishes as more excited states are occupied~\cite{thomas2017}. The efficiency enhancement is most robust to increasing temperature for long-range interactions, see Fig.~\ref{fig4}(b). At a given temperature $\beta\gtrsim 4\omega_0^{-1}$, the ratio $\sum_{i=2}^{2^N-1} n_i/n_1$ decreases as the range of interactions increases, see Fig.~\ref{fig4}(c). Here $n_i=e^{-\beta E_i}/Z(\beta,\omega)$ is the thermal occupation of energy level $i=0,...,2^{N}-1$, indexed in order of increasing energy $E_i$. Hence long-range interactions are most effective at suppressing fluctuations beyond the approximation Eq.~\eqref{2level}.

For sufficiently large $\beta_H\omega_0$, interactions degrade performance, see Fig.~\ref{fig4}(a),(b). Expanding the dimensionless free energy in powers of $\beta$, we obtain
\begin{equation}\label{ZhighT}
\ln Z=\ln Z_\infty+\frac{N\beta^2\omega^2}{8}+\frac{\beta^2g^2\sum_i(\Omega_i^\prime)^2}{8}+O(\beta^4),
\end{equation}
with $Z_\infty=2^N$ the infinite temperature partition function and $\Omega_i^\prime=\sqrt{\frac{1}{2}\sum_{j\ne i}J_{ij}^2}$. At order $\beta^2$, the free energy is indistinguishable from the mean-field free energy $\ln\operatorname{Tr}e^{\sum_i\beta(\omega\hat{\sigma}_z^{(i)}+\Omega_i^\prime\hat{\sigma}_x^{(i)})}$, in which case interactions degrade performance. The scaling $\ln (Z/Z_\infty)\propto \beta^2$ is clearly present for temperatures $\beta_H\lesssim \omega_0^{-1}$, see Fig.~\ref{fig4}(d). For temperatures $\beta\sim \omega_0^{-1}$, we expect an interplay between mean-field degradation and the enhancement~\eqref{2level}, which will govern the dependence of $g_\mathrm{max}$ on temperatures $\beta_H\lesssim 8\omega_0^{-1}$ (Fig.~\ref{fig4}(a)).

\section{Diabatic work extraction}\label{diabaticPerformance}

\begin{figure}
\includegraphics[trim=0cm 5cm 0cm 5cm,clip=true,width=0.5\textwidth]{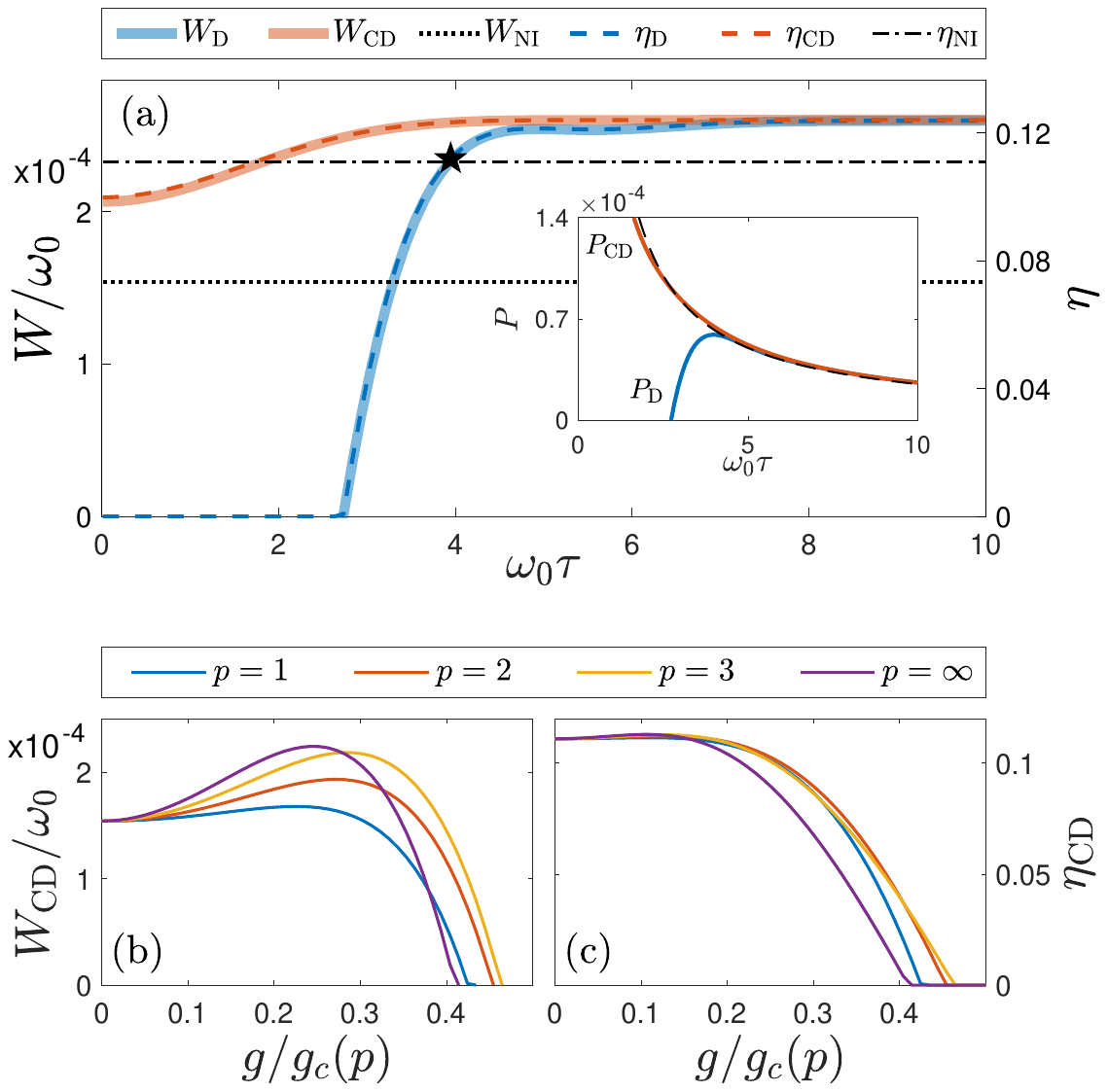}
\caption{\label{fig5} (a) Diabatic work output $W_\mathrm{D}$ and efficiency $\eta_\mathrm{D}$ for $p=\infty$ with weak interactions $g/g_c=0.2$. The maximum power output $P_\mathrm{D}$ (see inset) occurs at $\tau\approx 4\omega_0^{-1}$ (star), with performance rapidly declining for smaller $\tau$. The approximate counterdiabatic driving (Eq.~\eqref{cd}) results in work output ($W_\mathrm{CD}$) and efficiency ($\eta_\mathrm{CD}$) close to the adiabatic performance even for rapid engine cycles. Inset: The power output with counterdiabatic driving $P_\mathrm{CD}$ (red line) grows $\propto \tau^{-1}$ (black dashed line). The blue line is $P_\mathrm{D}$. (b),(c) The effectiveness of Eq.~\eqref{cd} diminishes for larger $g/g_c(p)$ or smaller $p$ (results for $\tau=\omega_0^{-1}$; for $p=\infty$, we use the exact $\chi_{ij}$, whereas for $p=1,2,3$, we set $\chi_{ij}=1$). All results are for $N=10$ and $r=r_\mathrm{NI}^\mathrm{max}$.}
\end{figure}

For diabatic (finite-time) work extraction, interactions generally degrade engine performance due to ``quantum friction''~\cite{kosloff2002,feldmann2003,plastina2014}. This friction arises when the interaction component of the Hamiltonian does not commute with the driving component, and hence the density matrix develops off-diagonal elements in the energy eigenbasis. The diabatic performance of a $p=\infty$ engine with weak interactions is shown in Fig.~\ref{fig5}(a). The peak power output occurs for a time step $\tau\approx 4\omega_0^{-1}$ (the precise value is dependent on $g$), at which point the efficiency is close to the adiabatic efficiency. For faster cycles, the performance rapidly decreases.

In principle, quantum friction can be mitigated completely using a counterdiabatic driving field $\hat{H}_\mathrm{cd}$~\cite{demirplak2003,berry2009}. In practice, exact counterdiabatic driving in a many-body system requires unrealistic interactions between all particles~\cite{delcampo2012,delcampo2013,deffner2014,saberi2014,campbell2015,passarelli2020,cepaite2022}, and approximate protocols are required. A powerful approximation method is to find $\hat{H}_\mathrm{cd}$ variationally by minimising the action $S=\operatorname{Tr}[G(\hat{H}_\mathrm{cd})^2]$, with $G(\hat{H}_\mathrm{cd})=\partial \hat{H}/\partial t+i[\hat{H}_{\mathrm{cd}},\hat{H}]$ and $\hat{H}_\mathrm{cd}$ expanded in some truncated set of operators~\cite{sels2017,kolodrubetz2017}. We use $\hat{H}_\mathrm{cd}=\sum_{ij(j\ne i)}C_{ij}\hat{\sigma}_x^{(i)}\hat{\sigma}_y^{(j)}$, which is the optimal counterdiabatic drive over all one-body and two-body operators (c.f.~\cite{hartmann2020}). For large $N$ in the paramagnetic phase, we obtain (see Appendix~\ref{cdderiv} for details),
\begin{equation}\label{cd}
\hat{H}_\mathrm{cd}=-\sum_{\substack{i,j=1\\(j\ne i)}}^N\frac{g\omega'(t)J_{ij}}{2\omega(t)^2}\chi_{ij}(t)\hat{\sigma}_x^{(i)}\hat{\sigma}_y^{(j)},
\end{equation}
with $\chi_{ij}(t)=1+O(g^2/\omega^2)$ given in Appendix~\ref{cdderiv}. The work protocols $f(t)$ and $f(\tau-t)$ satisfy $f'(0)=f'(\tau)=0$, and hence the net power transferred to the counterdiabatic drive field is zero. This can be shown explicitly by integrating by parts the instantaneous power $\langle\partial(\hat{H}+\hat{H}_\mathrm{cd})/\partial t\rangle$ and noting that only the boundary term remains.

For nearest neighbour interactions, $\chi_{ij}=1/(1+g^2/\omega(t)^2)$ and Eq.~\eqref{cd} drastically improves the diabatic engine operation for $g\lesssim 0.3g_c$, see Fig.~\ref{fig5}. For rapid cycles, the work output approaches a constant with little cost in efficiency, and hence the power output increases as $\tau^{-1}$, see inset to Fig.~\ref{fig5}(a). In practice, the time scale of the thermalization steps will limit the engine to finite power~\cite{curzon1975,geva1992}. Note $\eta\propto W$ irrespective of counterdiabatic driving (Fig.~\ref{fig5}(a)), hence $Q_H$ depends only weakly on $\tau$.

For increasing $g/g_c$ there is a trade-off in the performance gained from interactions and the performance lost from quantum friction, with peak performance occurring for $g/g_c\approx 0.3$ for $\tau=\omega_0^{-1}$. Here, the work output from a chain with nearest neighbour interactions is about $50\%$ larger than the non-interacting chain and both show comparable efficiency, see Fig.~\ref{fig5}(b),(c). For $p=1,2,3$, $\chi_{ij}(t)$ is difficult to engineer since the interactions must be reconfigured at different times. To simplify, we expand to lowest order in $g/\omega$ and set $\chi_{ij}(t)=1$. While this is somewhat effective at mitigating diabatic degradation for weak interactions, the performance enhancement diminishes as the range of interactions increases. Hence a chain with $p=1$, $g/g_c\lesssim 0.3$ and $\tau=\omega_0^{-1}$ has approximately the same performance as a non-interacting chain. Interestingly, we find that Eq.~\eqref{cd} is most effective for $\beta_H\lesssim 10\omega_0^{-1}$, with reduced performance for colder temperatures. This may be due to thermal fluctuations countering quantum friction~\cite{feldmann2006}.

\section{Conclusions}\label{conclusion}

We have shown that an engine of interacting spins outperforms a non-interacting engine in the paramagnetic phase for low temperatures and adiabatic operation, due to a lowering of the first excited state energy gap. The enhancement in work output is particular pronounced, with $W/W_\mathrm{NI}$ increasing exponentially with increasing interactions. The efficiency enhancement is largest for long-range interactions, which suppress occupation of energy levels beyond the first excited state. A performance enhancement due to long-range interactions has also been identified in Kitaev chains~\cite{wang2020,solfanelli2023}. For diabatic engine operation, quantum friction degrades performance. We have presented one counterdiabatic method that mitigates friction for weak interactions, however other methods could be explored~\cite{feldmann2006,delcampo2012,sgroi2021,cepaite2022,rolandi2023,deffner2020,cavina2021,soriani2022,erdman2023}. Modulating the phase and detuning of the drive profile may better isolate the two lowest energy eigenstates~\cite{facchinetti2016,facchinetti2018,williamson2020a,williamson2020b,cidrim2020}, limiting degradation due to thermal fluctuations and quantum friction. The low-temperature performance enhancement coincides with many-body quantum effects due to bipartite entanglement arising from the first excited state. A more thorough investigation of the entanglement properties of the thermal spin chain could reveal how entanglement changes for higher temperatures~\cite{gunlycke2001,osborne2002,fine2005,yu2009} or diabatic operation.

\section*{Acknowledgements}

We thank C. Woffinden and M. Edmonds for useful comments on the manuscript. This research was supported by the Australian Research Council Centre of Excellence for Engineered Quantum Systems (EQUS, CE170100009).

\appendix
\section{Determining $g_c(p)$}\label{gccalc}

\begin{figure}[H]
\includegraphics[trim=0cm 5cm 0cm 6cm,clip=true,width=0.5\textwidth]{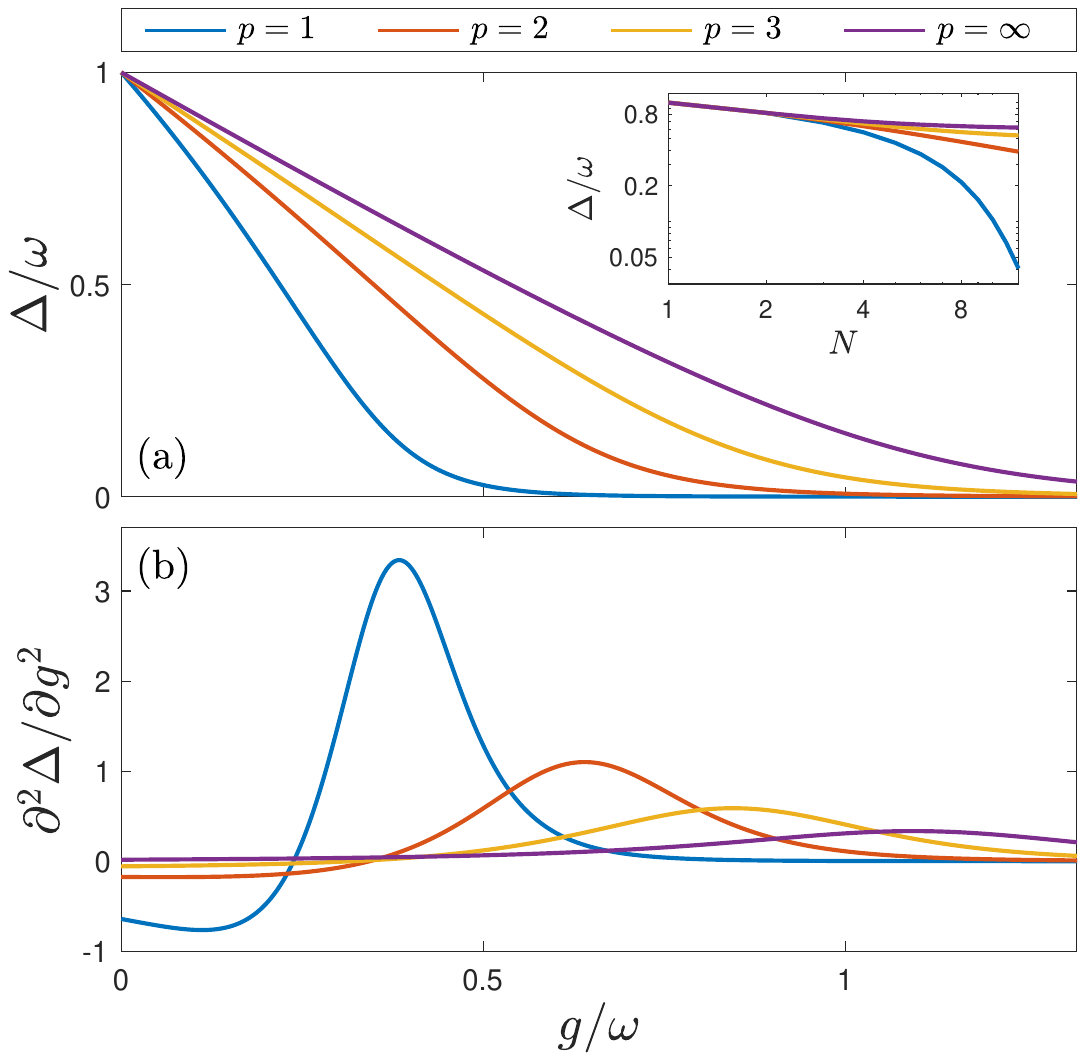}
\caption{\label{suppfig1} (a) Energy gap $\Delta$ between ground and first excited state for a 10-spin chain. The energy gap decreases as $\Delta=1-g/g_c(p)+O(g^2)$, where $g_c(p)$ is the critical interaction strength separating the paramagnetic and ferromagnetic phases. In the finite-sized system, we define $g_c(p)$ to be the point where $\partial^2\Delta/\partial g^2$, shown in (b), is a maximum. Inset in (a) shows the dependence of $\Delta$ on $N$ for $g=0.4\omega$.}
\end{figure}
In this appendix we explain how we determine the critical interaction strength $g_c$ separating the paramagnetic and ferromagnetic phases in a finite-sized spin chain. The energy gap $\Delta$ between the ground and first excited state is shown in Fig.~\ref{suppfig1}(a). The energy gap decreases approximately linearly with increasing $g/g_c(p)$. In an infinite system with nearest neighbour interactions, the exact energy gap is $\Delta=|\omega-g|$~\cite{pfeuty1970} and hence $\partial^2\Delta/\partial g^2=\delta(\omega-g)$. Finite-size effects regularize the divergence of $\partial^2\Delta/\partial g^2$, however we still observe a clear peak at a critical value of $g$, see Fig.~\ref{suppfig1}(b). We define $g_c(p)$ to be the value of $g$ corresponding to this peak. The critical value is close to $\omega$ for $p=\infty$ and decreases for decreasing $p$. The critical value will depend on $\omega$. We fix $g_c(p)$ by defining this to be the critical value at $\omega=\omega_0$.

\section{Bosonic spin-wave approximation}\label{perturb}
In this appendix we derive the bosonic spin-wave approximation for the Hamiltonian~\eqref{H}, which leads to the approximation Eq.~\eqref{2level}. We assume a large chain so that boundary effects can be ignored and hence we approximate the system as being translationally invariant (equivalently we can impose periodic boundary conditions). The spin operators can be converted to expressions in terms of bosonic operators via a Holstein-Primakoff transformation~\cite{holstein1940},
\begin{equation}
\begin{split}
\hat{\sigma}_x^{(i)}-i\hat{\sigma}_y^{(i)}\rightarrow &\left(\sqrt{1-\hat{a}_i^\dagger\hat{a}_i}\right)\hat{a}_i,\\
\hat{\sigma}_z^{(i)}\rightarrow &\frac{1}{2}-\hat{a}_i^\dagger\hat{a}_i.
\end{split}
\end{equation}
In the low excitation regime we expand the Hamiltonian~\eqref{H} to quadratic order in the bosonic operators $\hat{a}_i$, $\hat{a}_i^\dagger$. This gives~\cite{holstein1940,vogl2020,konig2021}
\begin{equation}
\hat{H}=-\frac{\omega N}{2}+\omega\sum_{i=1}^N\hat{a}_i^\dagger\hat{a}_i-\frac{g}{4}\sum_{\substack{i,j=1\\(j\ne i)}}^NJ_{ij}\left(\hat{a}_i+\hat{a}_i^\dagger\right)\left(\hat{a}_j+\hat{a}_j^\dagger\right).
\end{equation}
Fourier transforming and carrying out a Bogoliubov transformation gives~\cite{cevolani2016}
\begin{equation}
\hat{H}=\epsilon_0+\sum_{k=0}^{N-1} \left(\omega\operatorname{sign}\left(\omega-g\tilde{J}_k\right)\sqrt{1-\frac{2g}{\omega}\tilde{J}_k}\right)\hat{b}_k^\dagger \hat{b}_k,
\end{equation}
with $\epsilon_0$ the ground-state energy, $\hat{b}_k$ bosonic operators for the quasiparticle modes and
\begin{equation}\label{Jk}
\tilde{J}_k=\sum_{m=1}^{N-1} \frac{\cos\left(2\pi k m/N\right)}{m^p}=C_p^{N-1}\left(\frac{2\pi k}{N}\right).
\end{equation}
Here $C_p^N(\theta)=\sum_{m=1}^N\cos(m\theta)/m^p$ is the finite-$N$ generalized Clausen function~\cite{wu2010}. For nearest-neighbour interactions, $\tilde{J}_k=\cos (2\pi k/N)$ and hence $C_\infty^N(\theta)=\cos\theta$. From hereon we assume $\operatorname{sign}(\omega-g\tilde{J}_k)>0$.

The thermodynamic properties of the spin chain can be determined from the partition function,
\begin{equation}
Z=\prod_{k=0}^{N-1}\frac{1}{1-\exp\left(-\beta\omega\sqrt{1-\frac{2g}{\omega}\tilde{J}_k}\right)},
\end{equation}
with logarithm,
\begin{equation}
\begin{split}
\ln Z&=\sum_{k=0}^{N-1} \ln \left[\frac{1}{1-\exp\left(-\beta\omega\sqrt{1-\frac{2g}{\omega}\tilde{J}_k}\right)}\right],\\
&\approx\sum_{k=0}^{N-1}\exp\left(-\beta\omega\sqrt{1-\frac{2g}{\omega}\tilde{J}_k}\right).
\end{split}
\end{equation}
The latter approximation assumes low temperature. We have ignored the ground state energy, which adds an inconsequential constant to $\ln Z$. To proceed analytically, we assume small $g\tilde{J}_k/\omega$ and approximate $\sqrt{1-2g\tilde{J}_k/\omega}\approx 1-g\tilde{J}_k/\omega$. Hence
\begin{equation}\label{Zapprox}
\ln Z\approx e^{-\beta\omega}\sum_{k=0}^{N-1}\exp\left(\beta g\tilde{J}_k\right).
\end{equation}

We reserve a discussion of $p=1$ for later and for now assume $p>1$. For large $N$, $C_p^{N-1}(\theta)\rightarrow C_p(\theta)=\sum_{m=1}^\infty \cos(m\theta)/m^p$ and we can replace the sum over $k$ in Eq.~\eqref{Zapprox} by an integral
\begin{equation}
\sum_{k=0}^{N-1}\exp\left(\beta g\tilde{J}_k\right)\rightarrow \frac{N}{\pi}\int_0^\pi \exp\left(\beta g C_p(\theta)\right)\,d\theta\equiv N\mathcal{I}_p.
\end{equation}
For even integers $p$,
\begin{equation}
C_p(\theta)=-\frac{(-1)^{p/2}(2\pi)^p}{2p!}B_p\left(\frac{\theta}{2\pi}\right),
\end{equation}
with $B_p(x)$ the Bernoulli polynomials, which are polynomials of order $p$. We obtain the following results for $p=\infty$ and $p=2$,
\begin{equation}\label{p12}
\begin{split}
\mathcal{I}_\infty&=I_0(\beta g)\overset{\beta\rightarrow\infty}\sim \frac{e^{\beta g}}{\sqrt{2\pi \beta g}},\\
\mathcal{I}_2&=\frac{2 e^{\beta g\pi^2/6}}{\sqrt{\beta g\pi^2}}D\left(\frac{\sqrt{\beta g\pi^2}}{2}\right)\overset{\beta\rightarrow\infty}\sim \frac{e^{\beta g\zeta(2)}}{3\beta g\zeta(2)},
\end{split}
\end{equation}
where $I_0(x)$ is the modified Bessel function, $D(x)$ is Dawson's function (expressible in terms of the imaginary error function $\operatorname{erfi}(x)$ via $D(x)=(\sqrt{\pi}/2)e^{-x^2}\operatorname{erfi}(x)$) and $\zeta(2)=\pi^2/6$.

In general, for $p>3$ the integral $\int_0^1 \exp((m\beta gC_p(2\pi x)))\,dx$ can be approximated at low temperatures using the method of steepest descent,
\begin{equation}\label{pg3}
\mathcal{I}_p=\sqrt{\frac{1}{2\pi \beta gC_{p-2}(0)}}\exp(\beta gC_p(0)),\hspace{0.5cm}p>3,
\end{equation}
where $C_p(0)=\sum_{m=1}^\infty m^{-p}=\zeta(p)$. This does not work for $p=2$ since $C_2'(\theta)|_{\theta=0}\ne 0$. This reflects that the spectrum is linear rather than quadratic around the lowest energy state. Nor does it work for $p=3$, since $C_3^{''}(\theta)|_{\theta=0}$ diverges. We observe numerically that $\exp(\beta gC_3(\theta))$ is dominated by its small $\theta$ behaviour for large $\beta g$. Hence for $p=3$ we expand the Clausen function in a power series around the maximum $\theta=0$,
\begin{equation}
C_3(\theta)\approx \zeta(3)+\frac{1}{2}\theta^2\ln \theta-\frac{3}{4}\theta^2+O(\theta^4).
\end{equation}
This gives
\begin{equation}\label{p3}
\begin{split}
\mathcal{I}_3&\approx \frac{e^{\beta g\zeta(3)}}{\pi}\int_0^\pi\exp\left(-\frac{\beta g \theta^2}{4}\left(3-2\ln\theta\right)\right)\,d\theta,\\
&=\frac{e^{\beta g\zeta(3)}}{\pi}\int_0^\pi\exp\left(-\frac{\beta g \theta^2}{4}\left(3+\ln(\beta g)-\ln(\beta g\theta^2)\right)\right)\,d\theta,\\
&=\frac{e^{\beta g\zeta(3)}}{\pi\sqrt{\beta g}}\int_0^{\pi\sqrt{\beta g}}u^{u^2/2}\exp\left(-\frac{1}{4}u^2\left(3+\ln(\beta g)\right)\right)\,du,\\
&\overset{\beta\rightarrow\infty}\sim \frac{e^{\beta g\zeta(3)}}{\pi\sqrt{\beta g}}\int_0^{\pi\sqrt{\beta g}}\exp\left(-\frac{1}{4}u^2(3+\ln(\beta g))\right)\,du,\\
&=\frac{e^{\beta g\zeta(3)}}{\sqrt{\pi \beta g(3+\ln\beta g)}}\operatorname{erf}\left(\frac{\pi}{2}\sqrt{3\beta g+\beta g\ln\beta g}\right),\\
&\overset{\beta\rightarrow\infty}\sim \frac{e^{\beta g\zeta(3)}}{\sqrt{\pi \beta g(3+\ln\beta g})}.
\end{split}
\end{equation}
Hence we find a logarithmic correction $\ln \beta g$ to the partition function for $p=3$, which interpolates between $p=2$ and $p>3$. The asymptotic behaviour in the fourth line follows by replacing $u^{u^2/2}$ by $\lim_{u\rightarrow 0^+}u^{u^2/2}=1$ due to $\exp(-\frac{1}{4}u^2(\ln(\beta g)+3))$ being sharply peaked around the origin for large $\beta g$. The validity of this approximation was confirmed numerically. Equations~\eqref{p12}, \eqref{pg3} and \eqref{p3} give $\ln Z=N\mathcal{G}_p(\beta g)e^{-\beta \Delta}$ with $\Delta=\omega-g\zeta(p)$ and $\mathcal{G}_p(\beta g)$ arising from thermal fluctuations and dependent on $p$.

For $p=1$, Eq.~\eqref{Jk} at $k=0$ diverges with $N$ as $\tilde{J}_0\approx \ln N+\gamma$, with $\gamma$ the Euler-Mascheroni constant. We separate out this term in Eq.~\eqref{Zapprox}. For $k\ne 0$, Eq.~\eqref{Jk} converges and we have
\begin{equation}
\sum_{m=0}^\infty \frac{\cos(m\theta)}{m}=C_1(\theta)=-\ln |2\sin(\theta/2)|.
\end{equation}
Hence
\begin{equation}\label{freeenergyLR}
\ln Z\approx e^{-\beta \omega}e^{\beta g\gamma}N^{1+\beta g}+e^{-\beta \omega}\Lambda(\beta g)
\end{equation}
with
\begin{equation}\label{lambda}
\begin{split}
\Lambda(\beta g)=&2^{-\beta g}\sum_{k=1}^{N-1}\left(\sin \frac{\pi k}{N}\right)^{-\beta g},\\
=&2^{1-\beta g}\sum_{k=1}^{(N-1)/2}\left(\sin \frac{\pi k}{N}\right)^{-\beta g}
\end{split}
\end{equation}
arising from thermal fluctuations. We have assumed $N$ is odd in the second line in Eq.~\eqref{lambda}, however even $N$ will give the same final result below. The sum in $\Lambda(\beta g)$ is dominated by small $k$ terms, hence we can use the small-angle approximation $\sin(\theta)\approx\theta$,
\begin{equation}
\begin{split}
\Lambda(\beta g)&=2\left(\frac{N}{2\pi}\right)^{\beta g}\sum_{k=1}^{(N-1)/2}k^{-\beta g},\\
&\overset{N\gg 1}\approx 2\left(\frac{N}{2\pi}\right)^{\beta g}\zeta\left(\beta g\right).
\end{split}
\end{equation}
This term is smaller than the first term in Eq.~\eqref{freeenergyLR} by a factor $\sim N^{-1}(2\pi e^{\gamma})^{-\beta g}$. Hence thermal fluctuations are suppressed for low temperatures and large $N$, and we obtain,
\begin{equation}
\ln Z\approx Ne^{-\beta\Delta},
\end{equation}
with $\Delta\approx \omega-g(\ln N+\gamma)$. The low-temperature work output within this approximation is (see Fig.~\ref{suppfig2}(a)),
\begin{equation}\label{Wapprox}
\begin{split}
W&\approx N\omega_0(r-1)\left(e^{-\beta_H\Delta}-e^{-\beta_C\Delta}\right),\\
&\approx \omega_0(r-1)e^{-\beta_H\omega_0}e^{\beta_Hg\gamma}N^{1+\beta_Hg}.
\end{split}
\end{equation}

Summarising, to lowest order in $g/\omega$ and for large $N$ and $\beta$, we obtain the dimensionless free energies,
\begin{equation}
\ln Z\approx \left\{\begin{array}{ll}\sqrt{\frac{1}{2\pi\beta g\zeta(p-2)}}e^{-\beta\Delta},&\hspace{0.5cm}p>3,\\
\sqrt{\frac{1}{\pi\beta g(3+\ln\beta g)}}e^{-\beta\Delta},&\hspace{0.5cm}p=3,\\
\frac{1}{3\beta g\zeta(2)}e^{-\beta\Delta},&\hspace{0.5cm}p=2,\\
e^{-\beta\Delta},&\hspace{0.5cm}p=1,\end{array}\right.
\end{equation}
with $\Delta(\omega)=\omega-\omega_0 g/g_c$.

\section{Calculation of entanglement}\label{entangle}
In this appendix we calculate the low-temperature entanglement of the spin-chain. Within the bosonic approximation in Appendix~\ref{perturb}, the first excited state is a state with one excitation uniformly spread across all spins,
\begin{equation}
\ket{1}=\frac{1}{\sqrt{N}}\sum_{i=1}^N\hat{\sigma}_+^{(i)}\ket{0}.
\end{equation}
We consider a partition dividing the chain in half. The entanglement entropy of state $\ket{1}$ with respect to this partition is
\begin{equation}
\mathcal{S}=-\operatorname{Tr}_R(\rho_L\ln\rho_L),
\end{equation}
with $\rho_L=\operatorname{Tr}_L(\ket{1}\bra{1})$ and $\operatorname{Tr}_{R(L)}$ a partial trace over states in the right(left) half of the chain. It is straightforward to show that this gives $\mathcal{S}=\ln (N/2)$.

For low temperatures, we can qualitatively describe the system by the thermal state
\begin{equation}
\rho_1=\frac{1}{1+e^{-\beta\Delta}}\left(\ket{0}\bra{0}+e^{-\beta\Delta}\ket{1}\bra{1}\right),
\end{equation}
where $\Delta=\omega-\omega_0 g/g_c$ for small $g/\omega$. The Peres-Horodecki criterion states that a separable density matrix has positive partial trace~\cite{peres1996,horodecki1996}. Hence, if the density matrix does not have positive partial trace, the state is entangled (``non-PPT entanglement''). Due to the symmetry of the state, a sufficient condition for $\rho_1$ to exhibit non-PPT entanglement is $\operatorname{Tr}(\hat{M}\otimes\hat{M}\rho_1)<0$, with $\hat{M}$ any Hermitian operator acting on either the left ($\hat{M}\otimes$) or right ($\otimes \hat{M}$) partition of the chain~\cite{toth2009}. We choose $\hat{M}\otimes\hat{M}=\prod_{i=1}^N\hat{m}_i$ with
\begin{equation}
\hat{m}_i=\frac{1}{(1+\alpha)^{1/N}}\left[\hat{\sigma}_-^{(i)}\hat{\sigma}_+^{(i)}-\alpha\hat{\sigma}_+^{(i)}\hat{\sigma}_-^{(i)}\right],
\end{equation}
with $\alpha>0$ a free parameter. This gives
\begin{equation}
\operatorname{Tr}(\hat{M}\otimes\hat{M}\rho_1)=\frac{1-\alpha e^{-\beta\Delta}}{(1+\alpha)(1+e^{-\beta\Delta})}.
\end{equation}
We can make $\operatorname{Tr}(\hat{M}\otimes\hat{M}\rho_1)$ arbitrarily close to $-(1+e^{\beta\Delta})^{-1}$ by choosing $\alpha$ to be large ($\alpha\gg e^{\beta\Delta}$). This choice of $\hat{M}$ gives negative $\operatorname{Tr}(\hat{M}\otimes\hat{M}\rho_1)$ and hence the state exhibits non-PPT entanglement.

\section{Calculation of counterdiabatic drive}\label{cdderiv}
In this appendix we derive the approximate counterdiabatic drive Eq.~\eqref{cd}. We find the coefficients $C_{ij}$ in $\hat{H}_\mathrm{cd}=\sum_{ij(j\ne i)}C_{ij}\hat{\sigma}_x^{(i)}\hat{\sigma}_y^{(j)}$ by minimising the action
\begin{equation}
S=\operatorname{Tr}(G(\hat{H}_\mathrm{cd})^2),
\end{equation}
with $G=\frac{\partial \hat{H}}{\partial t}+i[\hat{H}_\mathrm{cd},\hat{H}]$~\cite{sels2017,kolodrubetz2017}. Hence we need to solve
\begin{equation}
\frac{\partial S}{\partial C_{ij}}=2\operatorname{Tr}\left(G\frac{\partial G}{\partial C_{ij}}\right)=0.
\end{equation}
For notational simplicity, we set $J_{ii}=C_{ii}=0$. We have
\begin{widetext}
\begin{equation}
\begin{split}
G&=-\omega'(t)\sum_p\hat{\sigma}_z^{(p)}-i\omega(t)\sum_{i,j,p}C_{ij}[\hat{\sigma}_x^{(i)}\hat{\sigma}_y^{(j)},\hat{\sigma}_z^{(p)}]-ig\sum_{i,j,p,q}C_{ij}J_{pq}\hat{\sigma}_x^{(i)}[\hat{\sigma}_y^{(j)},\hat{\sigma}_x^{(p)}\hat{\sigma}_x^{(q)}],\\
&=-\omega'(t)\sum_p\hat{\sigma}_z^{(p)}+\omega(t)\sum_{i,j,p}C_{ij}(\hat{\sigma}_x^{(i)}\hat{\sigma}_x^{(j)}\delta_{pj}-\hat{\sigma}_y^{(i)}\hat{\sigma}_y^{(j)}\delta_{pi})-g\sum_{i,j,p,q}C_{ij}J_{pq}\hat{\sigma}_x^{(i)}(\hat{\sigma}_z^{(p)}\hat{\sigma}_x^{(q)}\delta_{jp}+\hat{\sigma}_x^{(p)}\hat{\sigma}_z^{(q)}\delta_{jq}),\\
&=-\omega'(t)\sum_p\hat{\sigma}_z^{(p)}+\omega(t)\sum_{i,j}C_{ij}(\hat{\sigma}_x^{(i)}\hat{\sigma}_x^{(j)}-\hat{\sigma}_y^{(i)}\hat{\sigma}_y^{(j)})-2g\sum_{i,p,q}J_{pq}C_{ip}\hat{\sigma}_x^{(i)}\hat{\sigma}_x^{(q)}\hat{\sigma}_z^{(p)}.
\end{split}
\end{equation}
\end{widetext}
Hence
\begin{equation}
\frac{\partial G}{\partial C_{mn}}=\omega(t)(\hat{\sigma}_x^{(m)}\hat{\sigma}_x^{(n)}-\hat{\sigma}_y^{(m)}\hat{\sigma}_y^{(n)})-2g\sum_\ell J_{n\ell}\hat{\sigma}_x^{(m)}\hat{\sigma}_x^{(\ell)}\hat{\sigma}_z^{(n)}.
\end{equation}
We now want to calculate the trace of $G\partial G/\partial C_{mn}$ ($m\ne n$). The three terms in $G$ and the two terms in $\partial G/\partial C_{mn}$ gives a total of six terms. Three of these are zero, since the trace of terms with an odd number of spin-1/2 operators is zero. The three remaining terms are:

\begin{widetext}
\begin{equation}
\operatorname{Tr}\left[2g\omega'(t)\sum_p\hat{\sigma}_z^{(p)}\sum_{\ell}J_{n\ell}\hat{\sigma}_x^{(m)}\hat{\sigma}_x^{(\ell)}\hat{\sigma}_z^{(n)}\right]=2^N\frac{g\omega'(t)J_{mn}}{8}.
\end{equation}
\begin{equation}
\operatorname{Tr}\left[\omega(t)^2\sum_{p,q}C_{pq}(\hat{\sigma}_x^{(p)}\hat{\sigma}_x^{(q)}-\hat{\sigma}_y^{(p)}\hat{\sigma}_y^{(q)})(\hat{\sigma}_x^{(m)}\hat{\sigma}_x^{(n)}-\hat{\sigma}_y^{(m)}\hat{\sigma}_y^{(n)})\right]=2^N\frac{\omega(t)^2C_{mn}}{4}.
\end{equation}
\begin{equation}
\operatorname{Tr}\left[4g^2\sum_{p,q,r}J_{pq}C_{pr}\hat{\sigma}_x^{(r)}\hat{\sigma}_x^{(q)}\hat{\sigma}_z^{(p)}\sum_\ell J_{n\ell}\hat{\sigma}_x^{(m)}\hat{\sigma}_x^{(\ell)}\hat{\sigma}_z^{(n)}\right]=2^N \frac{g^2\left(J_{mn}(JC)_{nn}+\frac{1}{2}(J^2)_{nn}C_{mn}-J_{mn}^2C_{mn}\right)}{8}.
\end{equation}
\end{widetext}

Combining terms gives the coupled linear equations that determine $C_{mn}$,
\begin{widetext}
\begin{equation}
g\omega'(t)J_{mn}+2\omega(t)^2C_{mn}+g^2\left(J_{mn}(JC)_{nn}+\frac{1}{2}(J^2)_{nn}C_{mn}-J_{mn}^2C_{mn}\right)=0.
\end{equation}
\end{widetext}
This gives
\begin{equation}\label{Cmn}
C_{mn}=-\frac{g\omega'(t)J_{mn}+g^2J_{mn}(JC)_{nn}}{2\omega(t)^2+g^2\left[\frac{1}{2}(J^2)_{nn}-J_{mn}^2\right]}.
\end{equation}
Multiplying both sides by $g^2J_{nm}$ and summing over $m$ gives
\begin{equation}\label{Cnn}
g^2(JC)_{nn}=-(g\omega'(t)+g^2(JC)_{nn})\lambda_n,
\end{equation}
with
\begin{equation}
\lambda_n=\sum_m\frac{g^2J_{mn}^2}{2\omega(t)^2+g^2\left[\frac{1}{2}(J^2)_{nn}-J_{mn}^2\right]}.
\end{equation}
Equation~\eqref{Cnn} can be rearranged to give
\begin{equation}
g^2(JC)_{nn}=-\frac{g\omega'(t)\lambda_n}{1+\lambda_n}.
\end{equation}
Substituting this into Eq.~\eqref{Cmn} gives
\begin{equation}
C_{mn}=-\frac{g\omega'(t)\left(\frac{1}{1+\lambda_n}\right)J_{mn}}{2\omega(t)^2+g^2\left[\frac{1}{2}(J^2)_{nn}-J_{mn}^2\right]}.
\end{equation}
For an infinite chain with $J_{mn}=1/|m-n|^p$, $(J^2)_{nn}/2=\zeta(2p)$ and $\lambda_n=\lambda$ is independent of $n$. For nearest-neighbour interactions, $\lambda=g^2/\omega(t)^2$, and
\begin{equation}
C_{mn}=-\frac{g\omega'(t)}{2(\omega(t)^2+g^2)}\delta_{n,m+1}.
\end{equation}
Using Mathematica, we find that $\lambda$ can be evaluated analytically for selected values of $p$. For example, for $p=1$, $\lambda=1-x\cot x$ with $x=\pi g/\sqrt{2\omega(t)^2+\zeta(2)g^2}$.

\end{document}